\documentclass[namedreferences]{SolarPhysics}
\usepackage[optionalrh]{spr-sola-addons} 
\usepackage{graphicx}        
\usepackage{color}           
\usepackage{url}             

\newcommand{\aap}{    {\it Astron. Astrophys.}}

\newcommand{\aapr}{   {\it Astron. Astrophys. Rev.}}

\newcommand{\apj}{    {\it Astrophys. J.}}

\newcommand{\apss}{   {\it Astrophys. Spa. Sci.}}

\newcommand{\jgr}{    {\it J. Geophys. Res.}}

\newcommand{\solphys}{{\it Solar Phys.}}
\newcommand{\sovast}{ {\it Sov. Astronom.}}
\newcommand{\ssr}{    {\it Space Sci. Rev.}}
\newcommand{\araa}{    {\it Ann. Rev. Astron. Astrophys.}}

\begin{document}

\begin{article}

\begin{opening}

\title{Modelling of Nonthermal Microwave Emission From Twisted Magnetic Loops}

\author{I.\,N.~\surname{Sharykin}$^{1,2}$\sep
        A.\,A.~\surname{Kuznetsov}$^{2}$
       }
\runningauthor{Sharykin \& Kuznetsov}
\runningtitle{Modelling of Nonthermal Microwave Emission From Twisted Magnetic Loops}

   \institute{$^{1}$ Space Research Institute (IKI), Moscow, Russia\\
                     email: \url{ivan.sharykin@phystech.edu}\\
					$^{2}$ Institute of Solar-Terrestrial Physics, Irkutsk, Russia\\
                     email: \url{a_kuzn@iszf.irk.ru}}

\begin{abstract}

Microwave gyrosynchrotron radio emission generated by nonthermal electrons in twisted magnetic loops is modelled using the recently developed simulation tool {\it GX Simulator}. We consider isotropic and anisotropic pitch-angle distributions. The main scope of the work is to understand impact of the magnetic field twisted topology on resulted radio emission maps. We have found that nonthermal electrons inside twisted magnetic loops produce gyrosynchrotron radio emission with peculiar polarization distribution. The polarization sign inversion line is inclined relatively to the axis of the loop. Radio emission source is more compact in the case of less twisted loop, considering anisotropic pitch-angle distribution of nonthermal electrons.

\end{abstract}
\end{opening}

\section{Introduction}

The main source of information about the accelerated electrons produced during solar flares is observation of their hard X-ray (HXR) emission \cite{Kontar2011}. HXR bremsstrahlung emission is generated by nonthermal electrons with approximately power-law distribution interacting with protons of ambient solar plasma. Using some analytical models \cite{Brown1971,Syrovatskii1972}  one can extract information about the flux of the accelerated electrons and their power-law spectral index. Applying more sophisticated inversion technique of the X-ray spectrum \cite{Piana2007,Kontar2008} one can obtain spectra of the electron population numerically and without any ad-hoc assumptions about the spectral shape.

Another way to learn something about accelerated electrons is to study their radio emission \cite{Pick2008}. Basically, there are two types of radio emission associated with nonthermal electrons: coherent and incoherent. These two types correspond to different frequency ranges (slightly overlapping) and physical mechanisms. Coherent radio emission (e.g., type III bursts) up to $\sim 5$~GHz is produced by plasma waves (electron-cyclotron or Langmuir waves) excited by wave-particle interaction. Incoherent microwave emission ($\gtrsim 1$~GHz) is associated with gyrosynchrotron radiation of individual nonthermal electrons gyrating around magnetic field lines. In this paper we consider only the incoherent microwave emission of nonthermal electrons; this type of emission provides us with information about kinetics of nonthermal electrons in the coronal flaring loops as well as about the magnetic field strength and topology there. On the other hand, coherent emission is usually associated with electron beams passed through the corona and its study is more suitable for coronal plasma diagnostics.

Intensity of the HXR bremsstrahlung emission mostly depends on plasma density and nonthermal electron flux. Microwave radio emission brightness temperature and polarization degree also strongly depend on the nonthermal electrons pitch-angle distribution, magnetic field strength and its orientation relatively to the line-of-sight (LOS). Moreover, radio emission is not entirely optically thin. We should take into account radiative transfer to calculate the escaping flux of the microwave emission and its polarization. Direct computation of the gyrosynchrotron radio emissivity and absorbtion coefficient involves time-consuming numerical integration for all cyclotron harmonics \cite{Melrose1968,Ramaty1969}. However, there are some ways to simplify calculations. The simplest way to estimate parameters of nonthermal electrons with power-law distribution is to apply the formulae from the work of \opencite{Dulk1985} to observed gyrosynchrotron spectrum. This analytical approximation assumes uniform plasma density and magnetic field strength and isotropic pitch-angle distribution of nonthermal electrons and is valid only in the limited range of cyclotron harmonics and viewing angles relative to magnetic field. \opencite{Fleishman2010} introduced fast gyrosynchrotron codes, where the authors use some analytical approaches and numerical methods to evaluate the microwave radio spectra with high speed and good accuracy for different energy and pitch-angle distributions.

In reality, both the nonthermal electrons and thermal plasma fill the curved magnetic loops nonuniformly and, therefore, we need detailed three-dimensional modelling of microwave emission. The IDL-based widget tool {\it GX Simulator} \cite{Nita2015} allows us to do such 3D simulations. Using this interactive tool, one can import any magnetic field model and reconstruct magnetic flaring loop. Then we can define the spatial, pitch-angle and energetic distribution of the nonthermal electrons inside the magnetic loop. Numerical integration of the radiation transfer equation allows us to obtain the resulting radio brightness and polarization maps.

In the standard model of an eruptive two-ribbon flare \cite{Hirayama1974,Magara1996,Tsuneta1997} quasipotential magnetic loops are formed due to magnetic reconnection in the cusp under erupting plasmoid. These magnetic loops are filled with nonthermal electrons and we observe loop-like SXR sources connecting double HXR sources, which correspond to the loop footpoints. HXR observations made by RHESSI, Reuven Ramaty High Energy Solar Spectroscopic Imager \cite{Lin2002}, reveal a lot of loop-like HXR emission sources \cite{Battaglia2005,Jiang2006,Guo2012}. Nobeyama radioheliograph, NoRH \cite{Nakajima1995} observations also show us loop structures in the microwave range \cite{Kupriyanova2010,Morgachev2014}. Loop geometry of X-ray and microwave emission sources seems to be the usual observational manifestation of the flare energy release. The best way to understand the basic peculiarities of the radio brightness and polarization distribution along magnetic loop filled with nonthermal electrons is to consider potential magnetic field, as it possesses the simplest topology, describing flare loops. The work of \opencite{Kuznetsov2011} presents analysis of microwave emission from magnetic loops filled with nonthermal electrons. The authors discuss the effects of anisotropy and nonuniform distribution of nonthermal electrons along the loop. It was found that the effect of the electron anisotropy is the most pronounced near the footpoints and it also depends strongly on the loop orientation. Concentration of the emitting particles at the looptop results in a corresponding spatial shift of the radio brightness peak, thus reducing effects of the anisotropy. At frequencies around $10-20$ GHz, the spectrum is strongly dependent on the electron anisotropy, spatial distribution, and magnetic field nonuniformity.

Classical two-dimensional model of magnetic reconnection assume interaction of the opposite-polarity magnetic flux tubes at a null-point. But magnetic reconnection can occur in a magnetic configuration without null points as well. For example, twisted magnetic field flux ropes can experience internal magnetic reconnection \cite{Demoulin1996,Gordovskyy2011,Pinto2015}. In such case accelerated particles will be directly accelerated and injected into the loop volume \cite{Gordovskyy2011,Gordovskyy2012,Gordovskyy2013,Gordovskyy2014}. We know that orientation of the magnetic field in the flare region affects the spatial distribution of brightness and polarization of the microwave radio emission. Thus, one could suppose that twisted magnetic loop filled with nonthermal electrons will produce microwave emission with a peculiar (compared with potential loops) spatial distribution of brightness and polarization. These peculiarities probably depend on the twist degree. The main scope of this paper is to model the gyrosynchrotron radio emission from the twisted magnetic loop and understand influence of the twist degree on the spatial structure of the radio emission sources. We will investigate radio emission of nonthermal electrons with different pitch-angle distributions (isotropic and anisotropic) and also consider central and limb locations of the twisted magnetic loop. Any found peculiarities could provide us with a new tool of diagnostics of magnetic field topology in the region where nonthermal electrons propagate.

{\bf In this paper, we focus on theoretical aspects and numerical simulations; observational examples of (possibly) twisted magnetic loops have been omitted for clarity. The model magnetic configuration is described in Section \ref{loopmodels}. The radio simulations using {\it GX Simulator} are presented in Section \ref{gxmodel}. The results are discussed and conclusions are formulated in Section \ref{conclusion}.}

\section{Model of a twisted magnetic loop}\label{loopmodels}
To simulate twisted magnetic loop, we chose the analytical model of \opencite{Titov1999} developed to describe basic topology of the sheared twisted magnetic field in the active regions. The magnetic field is modelled by superposition of two opposite magnetic charges $q$, line current $I_0$ (connecting two charges and directed from positive to negative one) submerged under the photosphere, and a circular current $I$ in the plane perpendicular to the line current, which is an axis of the circle. Besides, circular current $I$ is distributed over some circular area $\pi a^2$, where $a$ is radius of the cross section of the tube. Thus, we actually consider current-carrying torus, which has coronal part corresponding to the loop. Finally, the value of the current $I$ is determined from the force balance condition. In the case of the thin current tube the magnetic configuration is very close to the force-free conditions, and that is why this model is often used as a test for non-linear force-free extrapolation algorithms \cite{Valori2010,Jiang2015}.

For our calculation we use the following values (very close to those used in the original work): $q=100$ T Mm$^{-2}$, $I_0=-7\times 10^{12}$ A, distance between the magnetic charges $L=100$ Mm, radius of the circular current $R=70$ Mm, and the depth of the line current $d=50$ Mm. The value of $a$ is determined by the twist number $N_t$, which corresponds to the number of the turns of the magnetic field line around torus. This parameter is of order of several units. We use the following formula \cite{Titov1999} to calculate $a$:
\begin{equation}
a = R\sqrt{N_t\frac{I_0}{I}}.
\end{equation}
Number of turns of the magnetic field lines around coronal part of the flux tube $N_c$ is given by
\begin{equation}
N_c = \frac{N_t}{\pi}\arccos\left(\frac{d}{R}\right).
\end{equation}

\begin{figure}
\centering
\includegraphics[width=1.0\linewidth]{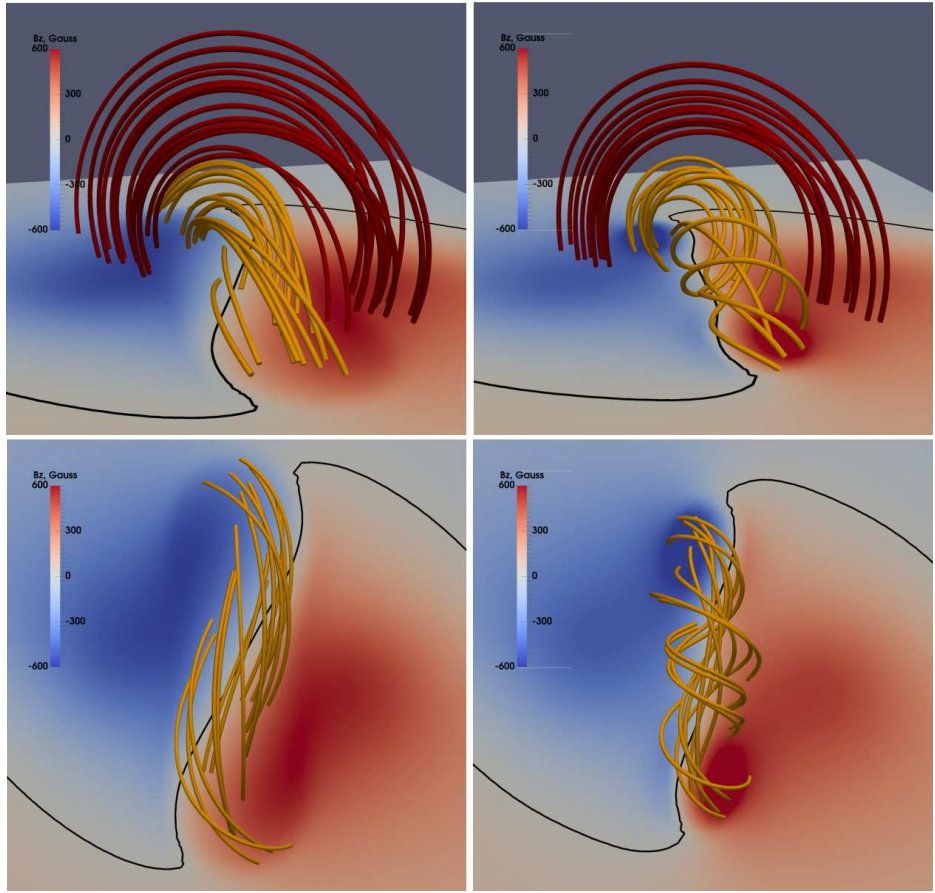}
\caption{The left and right columns correspond to the moderate and high twist cases accordingly. The top panels show top views of the synthetic photospheric magnetograms and reconstructed magnetic field lines. The bottom panels show stereoscopic views of the topology of the magnetic field lines.}
\label{TD_lines}
\end{figure}

\begin{figure}
\centering
\includegraphics[width=0.5\linewidth]{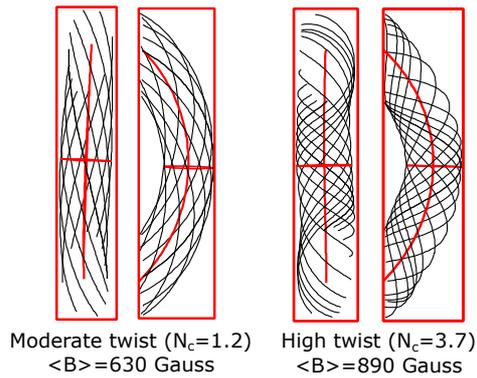}
\caption{Two variants of magnetic loops with different twist degrees calculated in {\it GX Simulator} and used for simulations in this paper.}
\label{TD_loops}
\end{figure}

In most calculations, we consider the cases of a moderate twist ($N_c\approx 1.2$, $N_t=5$) and a high twist ($N_c\approx 3.7$, $N_t=15$). For these values of $N_t$ the electric current density in the twisted magnetic loop is $j=I/(\pi a^2)\approx 2.5$ mA m$^{-2}$ for the moderate-twist case ($a\approx$~30~Mm) and $\approx 10$ mA m$^{-2}$ in the high-twist case ($a\approx$~15~Mm). After all calculations we rescaled the simulation box (to 40\% of its initial size) with preserving proportions and twist degree in order to obtain reasonable length of the loop of $\sim 20$~Mm which should be in accordance with typical observations. Top view of the synthetic photospheric magnetogram and reconstructed magnetic field lines for the above-mentioned twist numbers is shown in the top panels of Figure \ref{TD_lines}. Stereoscopic view of the magnetic field configuration is also presented in bottom panel of Figure \ref{TD_lines}. One can see that twisted magnetic configuration is surrounded by quasipotential sheared shell. We are interested only in the twisted magnetic field lines, so we will construct the radio-emitting magnetic loops only in the core of that region. Using the {\it GX Simulator} magnetic tube selection tool we defined approximately symmetric twisted magnetic loops, which are shown in Figure \ref{TD_loops}. One should note that the magnetic field strength does not varies much  along the loop; we have $B\approx 630$ G in the loop with moderate twist and $B\approx 890$ G in the highly-twisted loop. Thus distribution of the polarization and brightness of radio emission along the loop will be only affected by the topological peculiarities of the twisted magnetic field.

\section{Modelling of the nonthermal microwave emission from the twisted magnetic loops}\label{gxmodel}
\subsection{Parameters of the energetic electrons}
Nonthermal electrons are distributed along the loop uniformly with the number density of $10^8$~cm$^{-3}$. Their energetic spectrum is a power-law with the index of $\delta=3$, the low-energy cutoff $E_{\mathrm{low}}=10$ keV and the high-energy cutoff of $E_{\mathrm{high}}=10$ MeV. The background thermal plasma in the loop has the density of $n=5\times 10^9$~cm$^{-3}$ and the temperature of $T=20$ MK. We consider both the isotropic and anisotropic pitch-angle distributions of the nonthermal electrons. The anisotropic distribution has the Gaussian form:
\begin{equation}
g(\mu) = A\exp\left[-\frac{(\mu-\mu_0)^2}{\Delta\mu^2}\right],
\end{equation}
where $\mu_0=\cos\alpha_0$ characterizes the beam direction relatively to the magnetic field, $\Delta\mu$ is the width of the angular distribution, and $A$ is the normalization factor. We consider two types of anisotropy:

1) a pancake-like (or a symmetric loss-cone) distribution with $\mu_0=0$, which means mostly transversal propagation of the nonthermal electrons;

2) a beam-like distribution with $\mu_0=1$, when the nonthermal electrons propagate mostly along the magnetic field (the upward direction on the synthetic magnetograms in bottom panels of Figure \ref{TD_lines}).

Unless otherwise specified, we adopt the value of $\Delta\mu=0.15$ for both types of anisotropic distributions, which corresponds to a rather strong anisotropy.

To calculate the radio brightness maps, we use the fast gyrosynchrotron codes with radiation transfer \cite{Fleishman2010,Kuznetsov2011,Nita2015} implemented in the {\it GX Simulator}. To take into account the loop orientation relatively to the line-of-sight, we consider loop positions at the solar disk center and on the limb.

\begin{figure}
\centering
\includegraphics[width=0.7\linewidth]{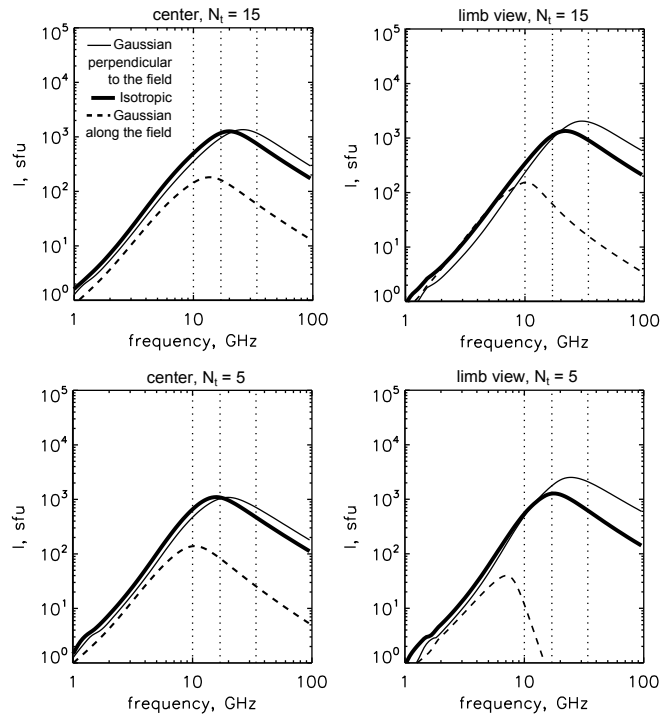}
\caption{The total (spatially integrated) spectra of the microwave gyrosynchrotron emission from the nonthermal electrons in the twisted magnetic loop. Top and bottom rows correspond to the cases of high and moderate twist of the loop (see Section \protect\ref{loopmodels}), respectively. Left and right columns show the spectra for the central and limb positions of the twisted loop. Spectra are calculated for the isotropic nonthermal electron distribution (thick line), anisotropic pancake-like distribution (thin line) and anisotropic beam-like distribution (dashed line); the pitch-angle distribution width for anisotropic distributions is $\Delta\mu=0.15$.}
\label{GS_spec}
\end{figure}

\subsection{Magnetic twist and electron anisotropy: general effects}
In Figure \ref{GS_spec} we show simulated spectra of the total (spatially integrated) emission. For central position of the loop the difference between isotropic and pancake-like cases is small: we only have a slight shift of the spectral peak and a bit enhanced radio emission at high frequencies for the pancake-like distribution. When the loop is at the solar limb, the anisotropy effect is more pronounced: the shift of the spectral peak is larger and the high-frequency emission is a few times more intense for the pancake-like electron distribution. However, in real flares these effects are expected to be not very significant, since the time-varying nonuniform distribution of the plasma density, magnetic field and parameters of the nonthermal electrons inside the loop will wipe out the spectral peculiarities connected with the twist of the loop. The nonthermal electrons with beam-like distribution produce the weakest emission at high frequencies comparing with the pancake-like and isotropic distributions.

In Figure \ref{GS_spec} we also indicate three frequencies (marked by dotted vertical lines): 10, 17 and 34 GHz. These different frequencies were selected to calculate the Stokes $I$ and $V$ maps and to compare the optically thin (34~GHz) and thick (10~GHz) cases. Radio emission at  17~GHz corresponds to the intermediate case with the optical depth of $\tau\sim 1$. The frequencies of 17 and 34~GHz are also the working frequencies of the Nobeyama Radioheliograph.

\begin{figure}
\centering
\includegraphics[width=1.0\linewidth]{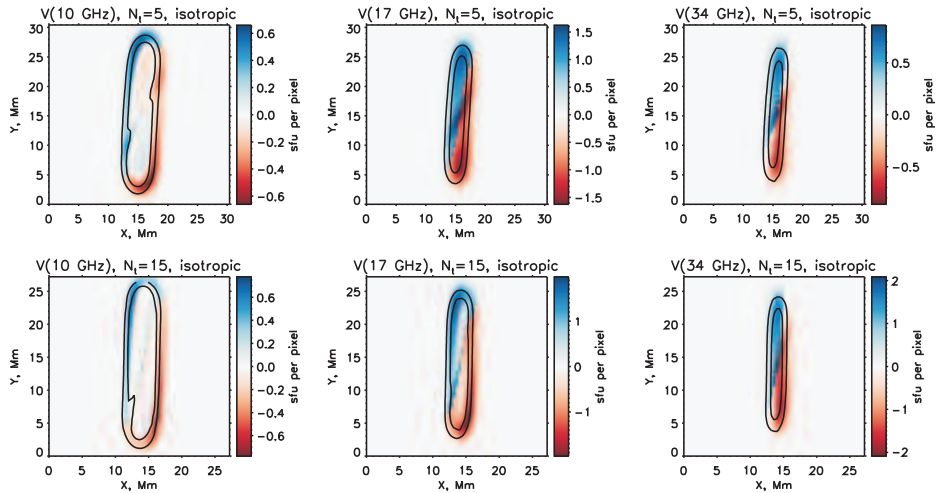}
\caption{Simulated radio maps for the magnetic loop at the solar disk center, for different frequencies, and loop twist degrees. Top three panels correspond to the strong twist while the bottom ones show the case of moderate twist. The pitch angle distribution of the nonthermal electrons is isotropic. Blue-red images show Stokes V. Black thick contours correspond to Stokes $I$ (at 30 and 70\% levels).}
\label{center_isotrop}
\end{figure}

Simulated Stokes $I$ and $V$ maps at the disk center for the isotropic nonthermal electron distribution are shown in Figure \ref{center_isotrop} (moderate and high twist). The observed width of the loop is maximal for the optically thick case at the frequency of 10~GHz and decreases towards higher frequencies and lower optical depths. The main peculiarity of the simulated maps is that the inversion of the polarization sign occurs near the inclined line crossing the loop. The angle of the inclination does not much depend on the twist degree. This orientation of the polarization sign inversion line (PSIL) is natural for a twisted magnetic loop. In the case of potential loop without any twist the PSIL will be approximately (depending on a loop orientation) perpendicular to the line connecting loop footpoints \cite{Kuznetsov2011}. Change of the circular polarization sign is associated with the change of the electron gyromotion direction (from clockwise to counter clockwise and vice versa) in the picture plane. In the twisted loop the electrons attached to different magnetic lines experience change of the gyromotion direction relative to the observer in different places of the loop (at different distances from its apex); therefore we observe strong inclination of the PSIL relative to the axis of the twisted loop.

\begin{figure}
\centering
\includegraphics[width=1.0\linewidth]{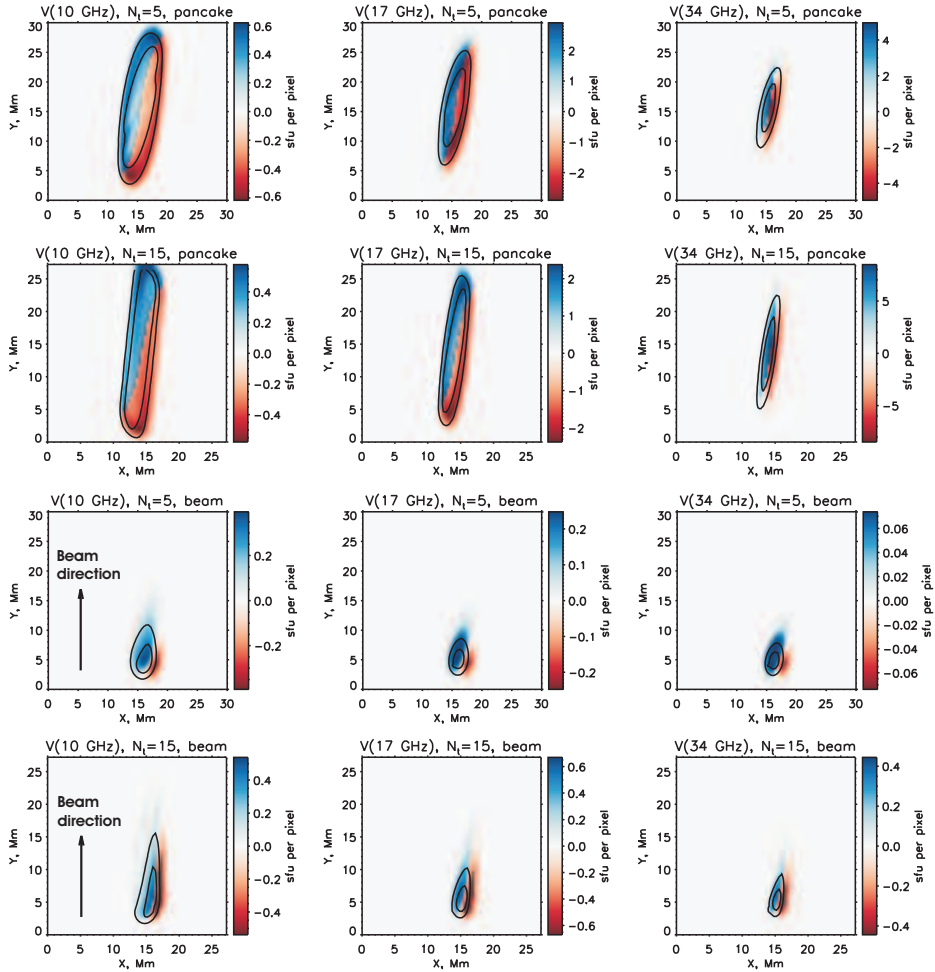}
\caption{Simulated radio maps for the magnetic loop at the solar disk center, for different frequencies, and loop twist degrees. Nonthermal electrons have the anisotropic pancake-like (top six images) and beam-like (bottom six images) pitch-angle distributions; $\Delta\mu=0.15$. Contours and color background have the same meanings as in Figure~\ref{center_isotrop}}
\label{center_anisotrop}
\end{figure}

Anisotopic nonthermal electrons with pancake-like pitch-angle distribution in the twisted loop produce microwave emission from a more compact source compared with the isotropic electrons (top six panels in Figure~\ref{center_anisotrop}). Like in the previous case, the PSIL is strongly inclined relative to the axis of the loop. Comparing with the isotropic case, the radio emission source itself is slightly inclined relative to the loop axis as well, but this effect can be negligible for different orientations of the loop relative to the observer. Another pronounced effect for the pancake-like distribution is a shrinkage of the radio emission source along the loop for lower twist: the radio emission source in the highly twisted loop is more elongated than in the case of the moderate twist. This effect is also connected with topology of the magnetic field. The gyrosynchrotron emission of the relativistic electron possesses strong directivity: it is collimated in the solid angle around the electron motion direction. The considered anisotropic nonthermal electron distribution in the loop with a moderate twist near the disk center produce emission mostly in the direction to the observer from the loop-top. In the highly twisted loop, gyrating electrons propagating along magnetic field lines have possibility to radiate radio emission effectively towards to the observer even from the footpoint region of the loop. However, a compact radio source can be formed also due to strong accumulation of the nonthermal electrons in the loop-top region (for example, due to the magnetic trapping), which is not considered in this work.

Polarization of the optically thin radio emission generated by nonthermal electrons with isotropic and pancake-like pitch-angle distributions corresponds to X-mode, while beam-like nonthermal electrons propagating along the magnetic field generate mostly O-mode radio waves \cite{Fleishman2003a,Kuznetsov2010}. In bottom six panels of Figure~\ref{center_anisotrop} we show Stokes $I$ and $V$ radio maps for the beam-like pitch-angle distribution. Direction of the beam coincides with the magnetic field direction, which is shown by an arrow in Figure \ref{center_anisotrop}. One can note that the PSIL intersects the footpoint region, where the radio emission is mostly localized. Radio emission source is more elongated in the case of the loop with higher twist degree, which is similar to the case of anisotropic pancake-like distribution. Thus a general property of the anisotropic nonthermal electrons distributions is shrinkage of the radio emission source in the magnetic loops with lower twist.

\begin{figure}
\centering
\includegraphics[width=1.0\linewidth]{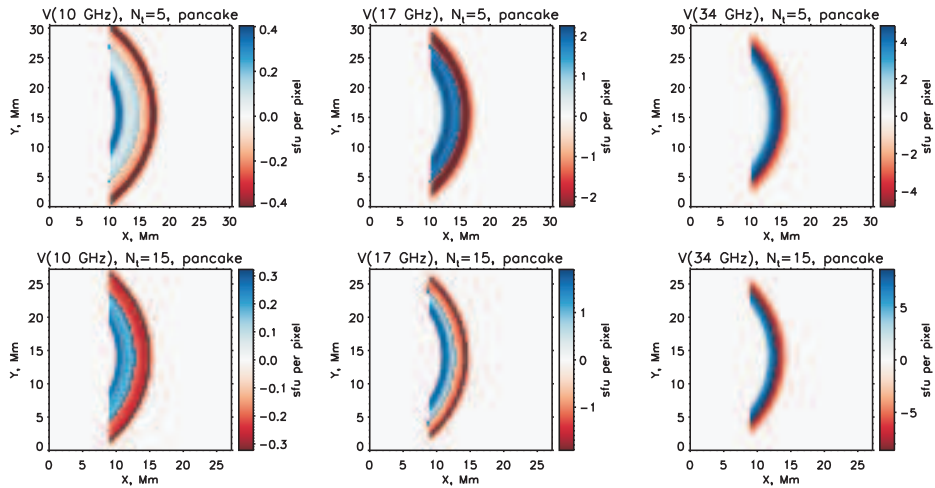}
\caption{Simulated radio maps for the magnetic loop on the solar limb with the moderate twist ($N_t=5$, top three images) and high twist ($N_t=15$, bottom three images). Nonthermal electrons have the anisotropic pancake-like pitch-angle distribution. The left, middle and right columns correspond to the frequencies of 10, 17 and 34 GHz, respectively. Blue-red images show the polarization (Stokes $V$) maps.}
\label{limb_pancake}
\end{figure}

Simulated radio maps for the limb position of the loop have similar qualitative peculiarities of emission source structure regardless of the twist degree and anisotropy of nonthermal electrons distribution. We show only one case for illustration in Figure \ref{limb_pancake} (circular polarization only), which corresponds to the anisotropic nonthermal electrons with pancake-like pitch-angle distribution. The PSIL divides the loop into two parts along its length and coincides with its axis. Distribution of the radio emission along the loop does not reveal any interesting features.

\begin{figure}
\centering
\includegraphics[width=1.0\linewidth]{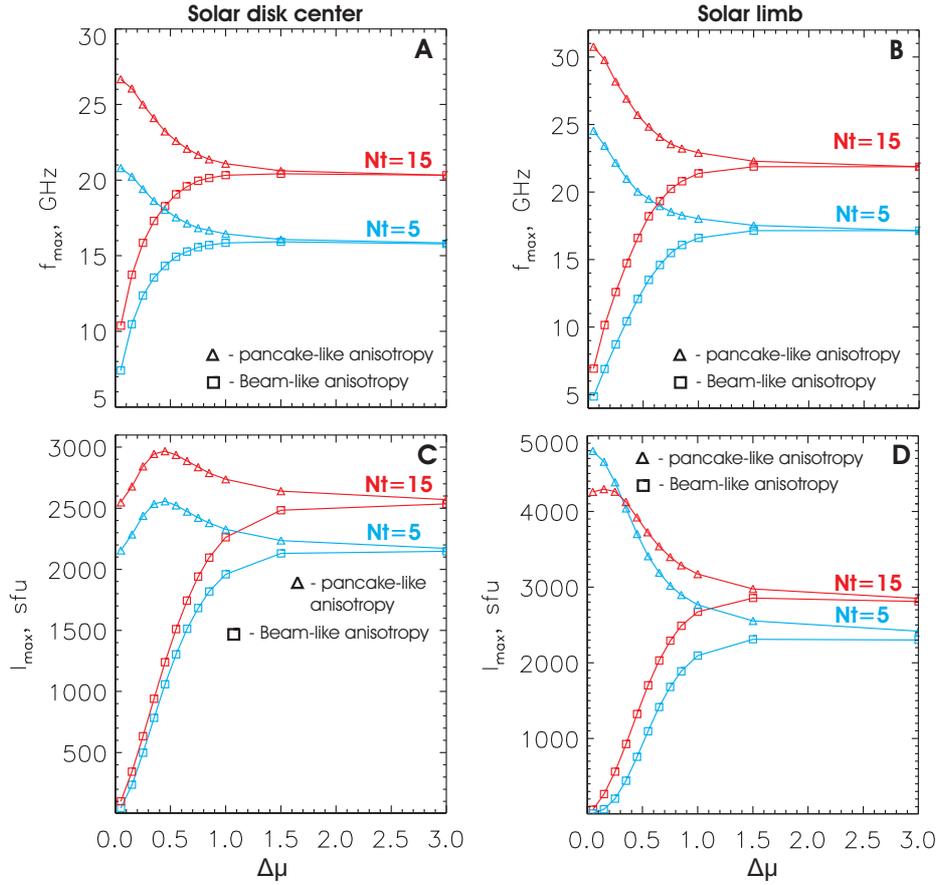}
\caption{The figure shows influence of nonthermal electrons anisotropy ($\Delta\mu$) on frequency ($f_{\max}$) of spectral maximum (panels A and B) and maximal intensity $I(f_{\max})$ (panels C and D). Cases of pancake-like (triangles) and beam-like (squares) anisotropy are considered. Panels A and C correspond to the central position of a loop on solar disk when other panels show the limb case. Red and blue colors mark high and moderate twist of a loop.}
\label{anisop_spec1}
\end{figure}

\subsection{Magnetic twist and electron anisotropy: parametric study}

We now consider quantitative dependencies of the radio emission on various parameters of its source. Influence of the nonthermal electrons anisotropy on the total (spatially integrated) microwave spectrum is demonstrated (for moderate and high twist) in Figure~\ref{anisop_spec1}; $\Delta\mu\ll 1$ corresponds to strong anisotropy (either beam-like or pancake-like), while $\Delta\mu\gg 1$ means a nearly isotropic distribution. One can note that the frequency ($f_{\max}$) of spectral maximum (panels A and B) decreases with increasing width ($\Delta\mu$) of the Gaussian pancake-like pitch-angle distribution. In the case of beam-like distribution we observe the opposite behaviour: increasing $f_{\max}$ corresponds to increasing $\Delta\mu$. Maximal spectral intensity (panels C and D) has tendency to decrease with increasing $\Delta\mu$ for pancake-like distribution and to increase for beam-like anisotropy of the nonthermal electrons. One can see that in the case of beam-like anisotropy radio intensity experiences significant changes due to variation of anisotropy degree for constant energetic spectrum and number density of nonthermal electrons. Such situation can arise in the case of isotropization of nonthermal electrons without significant dissipation: for example, magnetic field fluctuations can change direction of electrons motion without loss of their energy.

\begin{figure}
\centering
\includegraphics[width=0.5\linewidth]{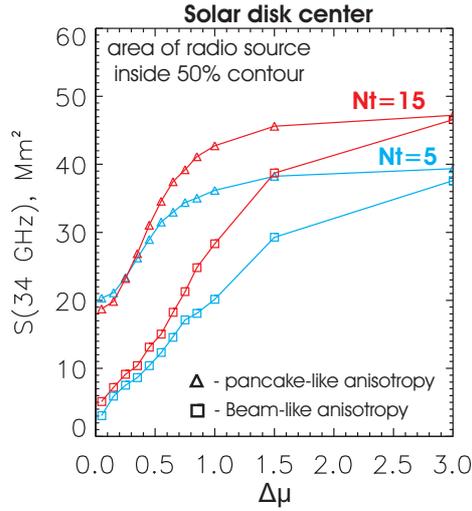}
\caption{The figure illustrates influence of nonthermal electrons anisotropy on microwave source size (area inside the 50\% contour) for the loop central position. Cases of pancake-like (triangles) and beam-like (squares) anisotropy are considered. Red and blue colors correspond to high and moderate twist of a loop.}
\label{anisop_spec2}
\end{figure}

Figure~\ref{anisop_spec2} shows how the nonthermal electrons anisotropy affects the microwave emission source size (which was determined as the area inside 50\% intensity contour); only the central position of the loop is considered. For smaller $\Delta\mu$ we have smaller sizes of radio sources for both types of anisotropy; this effect is more pronounced in the case of beam-like anisotropy. As expected, when $\Delta\mu$ increases, both the spectral peak frequency, maximum emission intensity and the emission source area for the beam-like and pancake-like distributions converge and approach the respective values for the isotropic electron distribution.

\begin{figure}
\centering
\includegraphics[width=0.8\linewidth]{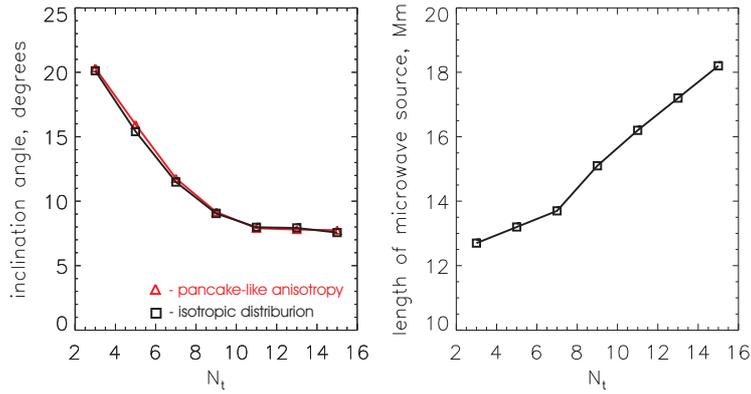}
\caption{Inclination angle of the PSIL relative to the loop axis is shown in the left panel as a function of twist degree($N_t$). Red color corresponds to the case of pancake-like anisotropy (with $\Delta\mu=0.15$), while black color marks isotropic nonthermal electrons. The right panel demonstrates influence of the twist degree on the length of the microwave source (at 50\% level) for the nonthermal electrons with pancake-like pitch-angle distribution ($\Delta\mu=0.15$).}
\label{Nt_trends}
\end{figure}

In the left panel of Figure~\ref{Nt_trends} we demonstrate how inclination angle between the PSIL and the loop axis (for the loop located at the disk center) depends on the twist degree $N_{t}$. The inclination angle is measured between the Y-axis and the line connecting two points corresponding to the intersections between the PSIL and the 50\% intensity contour. In the right panel of Fig.\,\ref{Nt_trends} we show the influence of the twist degree on length of the microwave source (measured at the 50\% level) for the case of nonthermal electrons with pancake-like pitch-angle distribution. One can note that larger twist corresponds to larger inclination of the PSIL to the loop axis; the resulting inclinations are almost the same both for isotropic and anisotropic nonthermal electrons. Larger twist also corresponds to longer microwave source for pancake-anisotropy of nonthermal electrons. We do not consider beam-like distribution in this parametric study as the PSIL is curvy (it is hard to determine the inclination to any direction) and the source shape is asymmetric. However, one can see that larger twist also corresponds to longer microwave source (Fig.\,\ref{center_anisotrop}). Length of the radio source in the isotropic case shows no significant dependence on the twist degree and, thus, we do not present it in the right panel of Fig.\,\ref{Nt_trends}.

\begin{figure}
\centering
\includegraphics[width=1.0\linewidth]{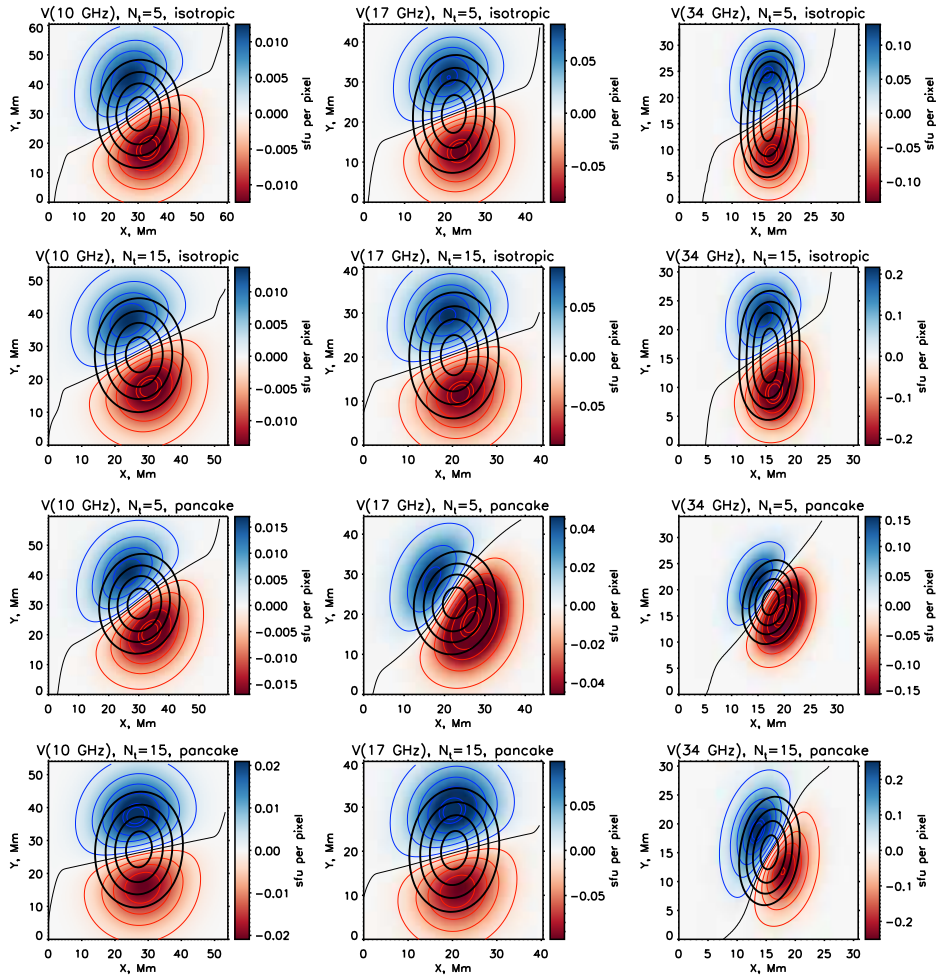}
\caption{Simulated radio maps for the magnetic loop at the solar disk center (similar to those in Figure \protect\ref{center_isotrop}) convolved with instrumental response functions, for different frequencies, loop twist degrees and electron pitch-angle distributions. Blue-red images and contours show Stokes $V$ maps (at 10, 30, 50, 70, 90 and 95\% levels); the black thin line marks the PSIL. Black thick contours correspond to Stokes $I$ (at 30, 50, 70 and 90\% levels).}
\label{NORH_sims}
\end{figure}

\subsection{Smoothed radio maps with reduced spatial resolution}
To obtain images similar to those observed by the Nobeyama Radioheliograph and other existing radio instruments, the simulated Stokes $I$ and $V$ radio maps are convolved with symmetric Gaussians (simulating point spread function, PSF) with the widths of 3.5~Mm ($\sim 5''$) for 34 GHz, 7~Mm ($\sim 10''$) for 17 GHz and 10~Mm ($\sim 17''$) for 10 GHz, respectively; the widths at 17 and 34 GHz correspond to the NoRH resolutions at these frequencies, while the width at 10 GHz is an extrapolation for a NoRH-sized instrument. We do not attempt to reproduce the real instrumental response function which depends on the heliograph base projection and hence varies according to day time and season; we are only interested in a qualitative picture and want to understand whether it is possible to detect any peculiarities of the gyrosynchrotron emission distribution in a twisted loop using the currently available observational data. The convolved radio maps for the loop located at the solar disk center are shown in Figure \ref{NORH_sims}; we consider here the isotropic and pancake-like pitch-angle distributions of nonthermal electrons.

After the convolution the resulting intensity per pixel is naturally reduced and the emission source sizes are increased. The PSIL for the convolved maps has a bit reduced inclination to the tube axis comparing with non-convolved radio maps. For anisotropic (pancake-like) nonthermal electrons, the size of the convolved emission source also depends on the twist degree of the loop: a more compact emission source is observed in the loop with moderate twist. The widths of the radio emission sources with different polarization signs are small comparing with the loop length. We do not present a figure showing the result of convolution of the limb images with a Gaussian because width of a Gaussian is larger than loop width and convolution leads simply to a corresponding increase of the visible loop width and to mutual compensation of the opposite polarizations.

The size of radio emission sources in the case of the beam-like pitch-angle distribution of nonthermal electrons is the smallest comparing with the isotropic and pancake-like distributions. We do not discuss the convolved images for this case, since we expect that all peculiarities of polarization will become undetectable in real observations.

Summarizing the results of the simulations, one can say that the most pronounced impact of the twisted magnetic field topology on the resulted spatial structure of the radio emission polarization is the inclination of the PSIL relatively to the loop axis. This effect is also pronounced in the smoothed radio maps. To find the flares where the PSIL is oriented relatively to the loop axis in the same manner as in our simulations, one should analyze the events with large-scale flare region ($\gtrsim 30$ Mm).

\section{Discussion and conclusions}\label{conclusion}
Obtained simulation results show that investigation of the spatial distribution of the radio emission polarization in the flare region can be used as twist diagnostics of the magnetic field where magnetized nonthermal electrons produce gyrosynchrotron emission. Moreover, it seems to be the only direct way to investigate topology of the magnetic fields in the region with accelerated electrons. Of course, twisting structures can be recognised by using high-resolution EUV and optical observations \cite{Wang2015}; however, those data (even together with X-ray observations) can only reveal local heating, which is not necessarily associated with propagating nonthermal electrons and cannot be considered as direct diagnostics of their activity.

In our modelling we use the simplest approach to describe the nonthermal electron population with analytical representation of their energetic and pitch-angle distributions. The modelling of the magnetic field is also simplified by using analytical derivations. To improve simulations of the radio emission from more or less realistic twisted magnetic fields, it would be better to use data-driven numerical 3D MHD simulations coupled with physics of electron acceleration and propagation. This task is very difficult technically and is a scope of future works. The main conclusions of the work can be summarized in a following way:
\begin{itemize}
\item Nonthermal electrons with isotropic pitch-angle distribution in the twisted loops produce gyrosynchrotron radio emission whose distribution along the loop does not show crucial dependence on the twist degree.
\item Nonthermal electrons with anisotropic pitch-angle distribution in the twis\-ted loop produce gyrosynchrotron radio emission from more compact source in the case of lower twist.
\item Inversion of the polarization sign of the radio emission, generated by nonthermal electrons in the twisted loop located in the center of the solar disk, has form of the line inclined relatively to the loop axis. Polarization of radio emission from twisted loop on the solar limb experiences change of its sign along its axis.
\end{itemize}

Definitely, to fully utilize the diagnostic potential of radio observations coupled with 3D simulations, we need new polarimetric multifrequency data with high spatial resolution. Very perspective instruments for these purposes are the Upgraded Siberian Solar Radio Telescope (uSSRT), Expanded Owens Valley Solar Array (EOVSA) and Mingantu Ultrawide Spectral Radioheliograph (MUSER).

As said above, in this paper we focus on numerical simulations and theoretical predictions. Some of the expected signatures of a twisted magnetic field (e.g., the specific orientation of the polarization inversion line) have been actually detected in the imaging observations of the Nobeyama Radioheliograph; these observations will be analyzed in detail in a forthcoming work.

\acknowledgements
This work was supported by the Russian Foundation for Basic Research (grant 15-32-51171). I.S. is grateful to the colleagues from the Institute of Solar-Terrestrial Physics for discussions and hospitality during staying at the ISTP.



\end{article}
\end{document}